# Coronal Magnetic Field Measurement from EUV Images made by the Solar Dynamics Observatory

Nat Gopalswamy[1], Nariaki Nitta[2], Sachiko Akiyama[1,3], Pertti Mäkelä[1,3], and Seiji Yashiro[1,3]

[1]NASA Goddard Space Flight Center, Greenbelt, Maryland

[2]Lockheed-Martin Solar and Astrophysics Laboratory, Palo Alto, California

[3]The Catholic University of America, Washington, DC

ABSTRACT

By measuring the geometrical properties of the coronal mass ejection (CME) flux rope and the leading shock observed on 2010 June 13 by the Solar Dynamics Observatory (SDO) mission's Atmospheric Imaging Assembly (AIA) we determine the Alfvén speed and the magnetic field strength in the inner corona at a heliocentric distance of ~ 1.4 Rs. The basic measurements are the shock standoff distance ($\Delta R$) ahead of the CME flux rope, the radius of curvature of the flux rope ($R_c$), and the shock speed. We first derive the Alfvénic Mach number (M) using the relationship, $\Delta R/R_c = 0.81[(\gamma-1) M^2 + 2]/[(\gamma+1)(M^2-1)]$, where $\gamma$ is the only parameter that needed to be assumed. For $\gamma = 4/3$, the Mach number declined from 3.7 to 1.5 indicating shock weakening within the field of view of the imager. The shock formation coincided with the appearance of a type II radio burst at a frequency of ~300 MHz (harmonic component), providing an independent confirmation of the shock. The shock compression ratio derived from the radio dynamic spectrum was found to be consistent with that derived from the theory of fast mode MHD shocks. From the measured shock speed and the derived Mach number, we found the Alfvén speed to increase from ~140 km/s to 460 km/s over the distance range 1.2 to 1.5 Rs. By deriving the upstream plasma density from the emission frequency of the associated type II radio burst, we determined the coronal magnetic field to be in the range 1.3 to 1.5 G. The derived magnetic field values are consistent with other estimates in a similar distance range. This work demonstrates that the EUV imagers, in the presence of radio dynamic spectra, can be used as coronal magnetometers.

## 1. Introduction

The idea of a shock standing ahead of a magnetic structure ejected from the Sun was proposed by Gold (1955, 1962), which was soon confirmed by space observations (Sonett et al., 1964). Extensive theory of the shock standoff distance has been developed for Earth's bow shock (see e.g., Bennett et al., 1997 and references therein). Russell and Mulligan (2002) applied the standoff distance derived by Farris and Russell (1994) to the case of a shock driven by an interplanetary flux rope (magnetic cloud) to explain the curvature of the flux rope derived from in situ observations. Gopalswamy and Yashiro (2011) demonstrated that the method can be applied to shocks driven by flux rope CMEs observed in SOHO/LASCO images at several solar radii from the Sun. White-light shock structures are observed for relatively fast CMEs (Gopalswamy et al., 2009, Ontiveros & Vourlidas, 2009). In this paper, we apply the standoff distance technique to a CME-driven shock observed on 2010 June 13 in the EUV images obtained by the Atmospheric Imaging Assembly (AIA, Lemen et al. 2011) on board the Solar Dynamics Observatory (SDO, Schwer et al. 2002). Gopalswamy and Yashiro (2011) had computed the ambient density in the corona using the polarized brightness image of the corona, but here we derived the density from the accompanying type II radio burst observed by the Hiraiso Radio Spectrograph (HiRAS) in Japan (Kondo et al., 1997).

## 2. Observations and Analysis

The 2010 June 13 CME was observed in several EUV wavelengths by SDO/AIA. Here, we use the AIA 193 Å images. The CME was described as a bubble by Patsourakos et al. (2010), who treated the CME as a sphere and reported on the early kinematics of the CME. Here we focus on the overlying shock structure that surrounds the flux rope and its standoff distance. Physical properties of the shock structure have also been reported by Ma et al. (2011). Kozarev et al. (2011) also identified the shock wave and reported on the implications for a possible solar energetic particle event. Unlike these authors, we focus on the standoff distance of the shock as a function of the heliocentric distance to derive the upstream magnetic field.

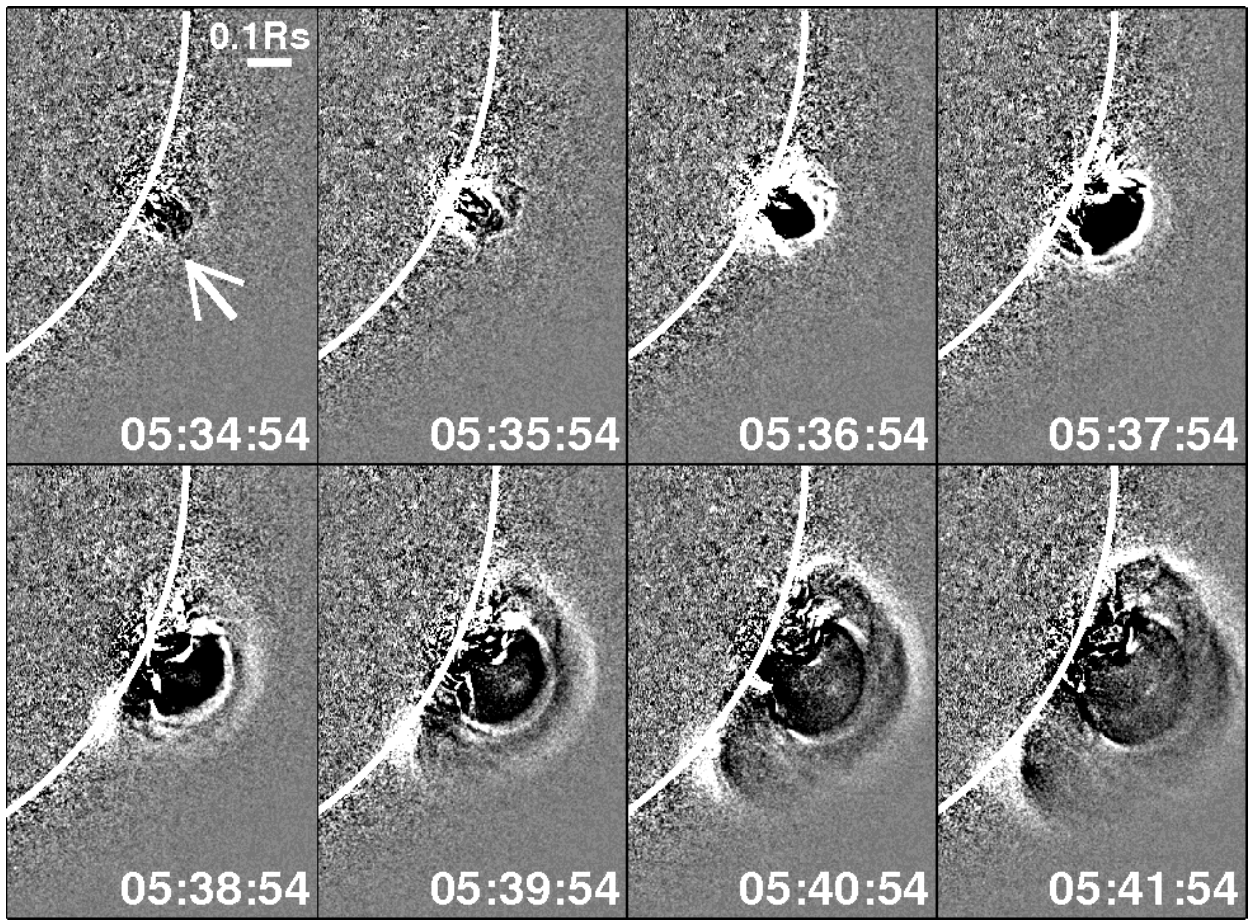

Figure 1. Snapshots of the CME from SDO/AIA 193 Å difference images. A small section of the images from the southwest corner is shown (the image scale is indicated in the first frame). The arrow points to the early stage of the CME. The scale of the images is also indicated on the first frame in terms of a solar radius (Rs). The optical limb is also drawn. A full cadence (12 s) movie of the event is included as an electronic supplement.

**2.1 EUV Observations**

Since February 2010, the AIA has been taking full-disk coronal images of the Sun in seven EUV channels representing a wide range of temperature. The pixel resolution is 0.6" and the basic cadence is 12s (see Lemen et al., 2011 for more details of the instrument). Here we sample the images in 193 Å every minute starting from the time the first signatures of the eruption were observed. The eruption occurred from active region NOAA 1079 from the southwest limb (S25W84) in association with an M1.0 flare. Figure 1 shows running difference images, in which we see the evolution of the CME as a series of snapshots from the very early stage at 05:34:54 UT to 05:41:54 UT. The shock structure becomes clear in the 05:37:54 UT image and is visible throughout the observation.

The shock structure connects to the EUV wave feature moving away from the eruption site, more clearly seen at later times. This is consistent with the interpretation that the EUV wave is a fast mode shock surrounding the CME flux rope (see e.g., Veronig et al., 2010). We identify the inner circular feature as the flux rope that drives the shock. There is a sudden change in the shape of the flux rope indicating rapid expansion in the lateral direction between 05:36:54 and 05:37:54 UT. Interestingly this is the time the shock became distinctly visible. Full-cadence data show that a diffuse structure formed ahead of the flux rope at 05:36:54 UT. This coincided with the onset of the metric type II radio burst (see below). The overall position angle extent (including the shock) was $\sim 37^{o}$ by the time the CME left the SDO/AIA field of view (FOV). The flux rope itself subtended an angle of only $\sim 17^{o}$. The CME first appeared in the LASCO/C2 FOV at 06:06 UT, when the leading edge was already at 2.57 Rs. The speed within the LASCO FOV is only 320 km/s and the CME has a position angle extent of $\sim 33^{o}$. Clearly, the flux rope seems to have decelerated significantly by the time it reached the outer corona. The position angle extent suggests that the LASCO CME must be the expanded flux rope. The CME does not show the diffuse shock structure in the LASCO FOV, suggesting that the shock might have weakened and dissipated by the time it reached the LASCO FOV. Therefore the speed must have peaked somewhere between 1.5 and 2.57 Rs. Without the SDO observations, the connection between the type II burst, the shock and the flux rope would not have been revealed.

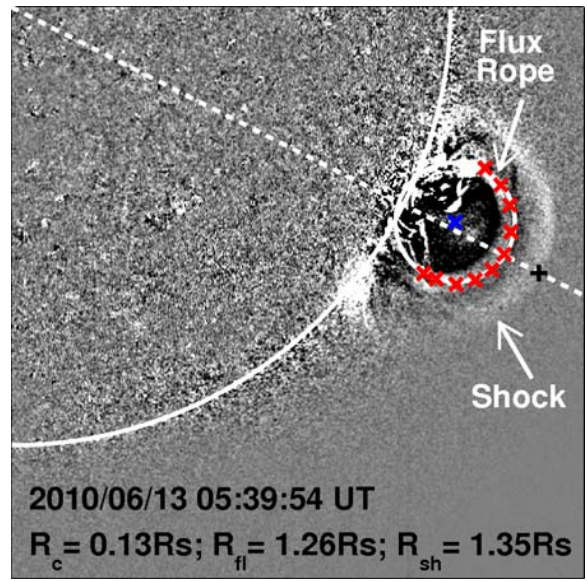

Figure 2. SDO/AIA difference image at 193 Å showing the flux rope and the shock structure surrounding it. The heliocentric distances of the shock ($R_{sh}$, marked by the '+' symbol) and the flux rope ($R_{fl}$) are shown. The red crosses are the points on the flux rope used for fitting the circle. The blue cross marks the center of the fitted circle. The radius of the circle fitted to the flux rope is the radius of curvature ($R_c$) of the flux rope.

The shock structure is better described in Fig. 2 with the projection of the flux rope on the sky plane fitted to a circle. We identify the radius of this circle as the minor radius of the flux rope and also the radius of curvature of the shock driver. The shock itself is not resolved, but the broad diffuse feature ahead of the flux rope is the shock sheath. The leading edge of the shock sheath is expected to be the shock. We identify the position of the leading edge of the sheath as the shock location. Thus the thickness of the sheath becomes the standoff distance of the shock ahead of the flux rope.

The heliocentric distance of the shock ($R_{sh}$) was measured at six instances from 05:36:54 UT to 05:42:54 UT, traveling from ~1.19 Rs to 1.46 Rs. The flux rope leading edge ($R_{fl}$) was observed from ~1.13 Rs to 1.38 Rs before its leading edge left the SDO/AIA field of view after 05:43:54 UT. The standoff distance (the difference between $R_{sh}$ and $R_{fl}$) steadily increased from ~0.02 Rs to 0.13 Rs when both the flux rope and the shock were observed. The flux rope radius is the same as the radius of curvature ($R_c$), which doubled during the flux rope transit through the SDO/AIA FOV. This is a limb event, so we are looking at the cross section of the flux rope, with its axis roughly perpendicular to the sky plane. Note that the radius of curvature of the flux rope is initially similar to half of the flux rope height above the limb, but rapidly increases at the time of shock formation. The eruption is from W84, so the projection effects are expected to me minimal: the true heights and speeds are expected to be larger only by < 0.6%, which is negligible.

Figure 3 shows the height-time plots of the shock, the flux rope and the radius of the flux rope. The average speeds of the flux rope and shock are 330 km/s and 644 km/s, respectively. The average expansion speed of the flux rope is ~160 km/s. The diameter of the flux rope increases with the same speed as the radial speed of the flux rope. In Figure 3(a), the height-time measurements are fitted with second order polynomials. Both the flux rope and the shock show increase in speed through the SDO FOV. The sudden increase in the flux rope radius between 05:36:54 UT and 05:37:54 UT is associated with the rapid expansion of the flux rope and coincides with the shock formation. A third-order polynomial fit to the flux rope radius shows a complex evolution compared to the leading edge of the flux rope and the shock. From the first three points, we estimate an average expansion speed of ~42 km/s, which jumps to 136 km/s by 05:37:54 UT around the time of shock formation. The radial speed of the CME flux rope is only ~245 km/s at the time of shock formation.

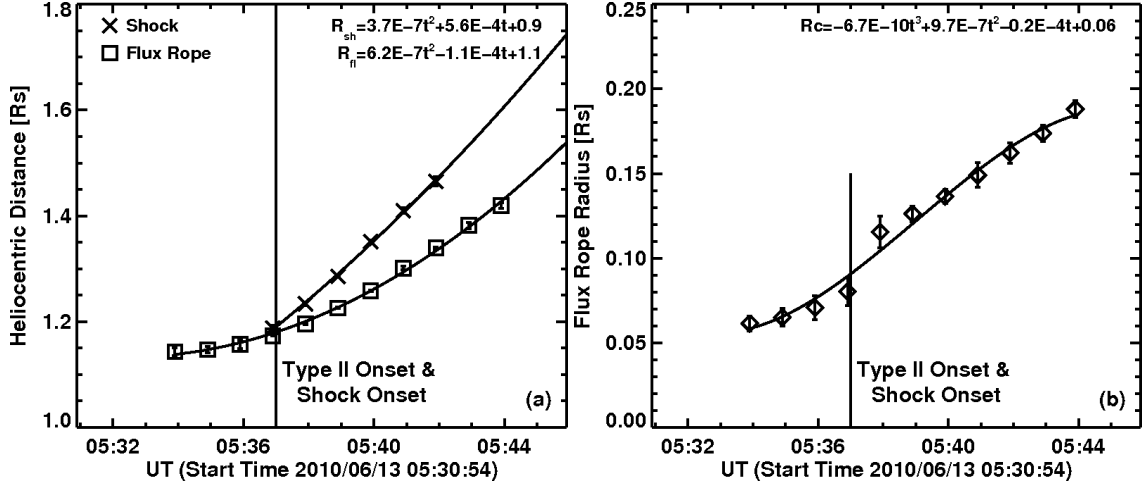

Figure 3. (a) height – time plot of the shock and flux rope with the quadratic fit to the measurements. One-minute difference images were used. The shock first appeared only at 05:36:54 UT, which coincided with the onset of the metric type II burst. The error bars on the heights give standard deviations of four independent measurements. The equations of the fitted curves for $R_{sh}$ (t) and $R_{fl}$ (t) are given on the plot, where t is the time in seconds from 05:30:54 UT. (b) Time evolution of the flux rope radius (also known as minor radius) obtained by fitting a circle to the flux rope cross section. The error bars represent the errors in fitting the circle. The solid curve is the third order polynomial fit to the data points, whose equation is shown on the plot: $R_c$ (t) with t in seconds from 05:30:54 UT.

## 2.2 Radio Observations

The shock formation observed in EUV remarkably coincided with the onset of a metric type II burst at 05:37:00 UT at the frequency of 300 MHz (harmonic) and 150 MHz (fundamental). Type II radio bursts are emitted at the local plasma frequency in the vicinity of the shock, so a 150 MHz implies a local plasma density of $2.8 \times 10^8 cm^{-3}$, which is consistent with the shock formation close to the solar surface (0.19 Rs above the surface). The radio dynamic spectrum from the Hiraiso Radio Observatory in Fig. 4 shows that the fundamental and harmonic components are visible, so there is no ambiguity in relating the emission frequency to the plasma frequency. The harmonic component is visible better than the fundamental component. The average drift rate of the type II burst is ~0.28 MHz/s, which is typical of metric type II bursts (see Gopalswamy et al., 2009). The harmonic component also shows band splitting, which indicates that the emission comes from behind and ahead of the shock (Smerd et al., 1974; Vrsnak et al., 2004; Cho et al., 2007). At the five times marked in Fig.4, the upper ($f_2$) and lower ($f_1$) bands are at frequencies (186, 156), (162, 128), (146, 118) (128, 104), (115, 90) MHz, indicating an average separation of ~28 MHz. The frequency ratio $f_2/f_1$ is directly related to the density jump across the shock because $f_1$ and $f_2$ correspond to emission ahead and behind the shock. The average value of $f_2/f_1 = 1.24$, which corresponds to a density compression of 1.54 across the shock.

To derive the magnetic field from the Alfvén speed, one needs the ambient plasma density at the shock nose. For the 2010 June 13 event, we can use the type II radio burst observations to get the plasma density because the emission takes place at the local plasma frequency ($f_p$) and its second harmonic ($2f_p$). Since $f_p = 9\times10^{-3}n^{1/2}$ MHz with the electron density n in $cm^{-3}$, one can obtain n from the dynamic spectrum by identifying $f_p$. The emission ahead of the shock is from the unshocked corona, so the emission occurs at a lower frequency ($f_1$) compared to the compressed downstream ($f_2 > f_1$). We need to use the lower frequency ($f_1$) to get the upstream plasma frequency ($f_{p1}$) and density ($n_2$): $f_{p1} = f_1/2$. Figure 4 shows that the band splitting is clear between 05:39:30 and 05:43:30 UT. Three of the SDO/AIA frames overlap with this interval: at 05:39:54,

05:40:54, and 05:41:54 UT, so we get $f_1$ = 128, 118, and 104 MHz giving the upstream densities ($n_2$) as 5.1, 4.3, and 3.3 in units of $10^7 cm^{-3}$ at these times.

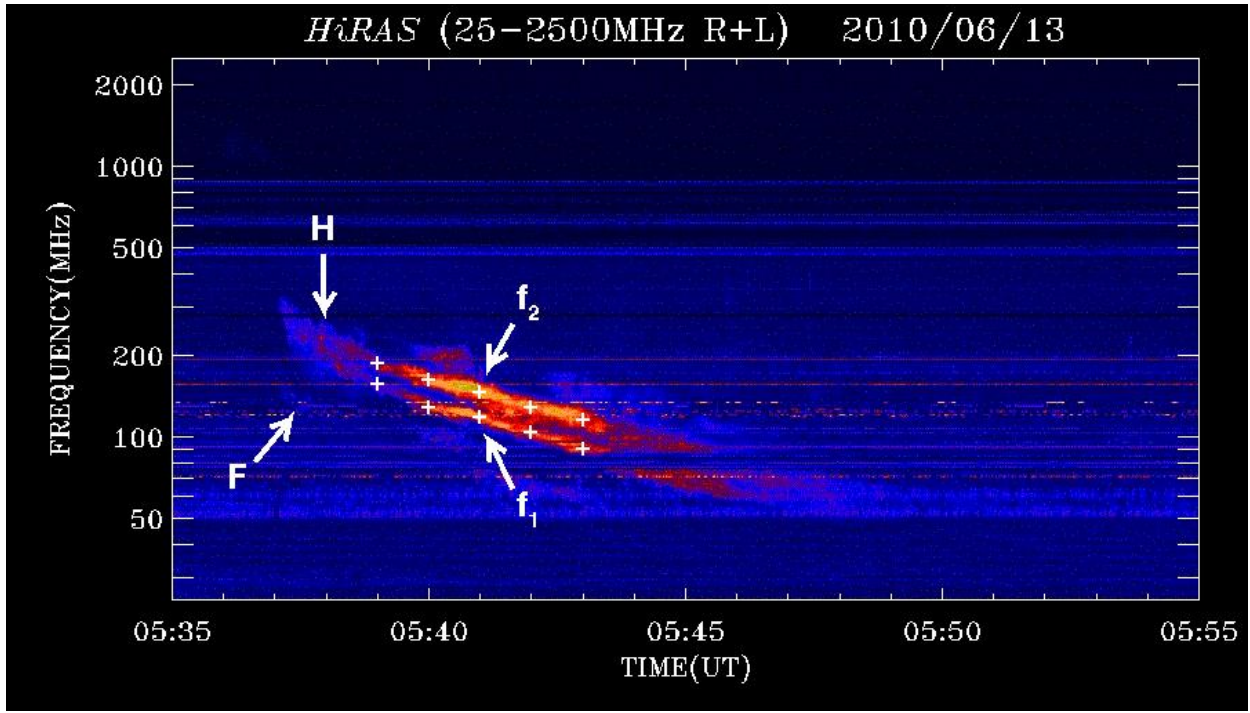

Figure 4. Radio dynamic spectrum from the Hiraiso Radio Spectrograph (HiRAS) showing the type II burst starting from ~05:37 UT until about 05:47 UT beyond which it is masked by the interference in the 50 – 60 MHz frequency range.  The band splitting is clear between 05:39:30 and 05:43:30 UT.

**2.3 Coronal Magnetic Field**

The standoff distance method to derive the coronal magnetic field involves measuring the leading edges of the CME flux rope ($R_{fl}$) and the leading shock ($R_{sh}$) at the nose and measuring the radius of curvature ($R_c$) of the flux rope fitted to a circle (see Gopalswamy and Yashiro, 2011).  The standoff distance $\Delta R = R_{sh} - R_{fl}$ is related to the shock Mach number (M) and the adiabatic exponent ($\gamma$) (Russell and Mulligan, 2002):

$$\Delta R/R_c = 0.81[(\gamma-1)M^2 + 2]/[(\gamma+1)(M^2-1)]. \quad \ldots(1)$$

Here we take the Mach number to be the Alfvén Mach number. In terms of the relative standoff distance $\delta = \Delta R/R_c$, the Mach number can be written as

$$M^2 = 1 + [1.24\delta - (\gamma-1)/(\gamma+1)]^{-1}. \tag{2}$$

Measuring $\delta$ and assuming $\gamma$, one can get the Alfvén Mach number. Note that the second term on the right hand side of Eq. (2) needs to be positive for a shock, which imposes a minimum value of $\delta = 0.115$ for $\gamma = 4/3$ and $0.202$ for $\gamma = 5/3$. In order to get the magnetic field strength (B), we need the upstream Alfvén speed ($V_A$) and the plasma density (n). Since $V_A = V_{sh}/M$, where the shock speed $V_{sh}$ is obtained from the time series of $R_{sh}$ measured in the EUV images. In the coronal region we are interested in, the solar wind speed is negligible compared to the shock speed. The magnetic field strength in the upstream medium is given by,

$$B = 4.59 \times 10^{-7} V_A n^{1/2} \text{ G}. \tag{3}$$

All we need is the upstream density, which can be derived directly from the emission frequency of the type II burst as explained in Section 2.2. For example, at 05:40 UT, the upper and lower bands of the harmonic component have frequencies $f_2=162$ and $f_1=128$ MHz, respectively. The local plasma frequencies are therefore, $f_{p2} = 81$ and $f_{p1} = 64$ MHz, respectively, corresponds to $n_1 = 5.1 \times 10^7 \text{cm}^{-3}$. At 05:39:54 UT, we measure $\Delta R = 9.1 \times 10^{-2}$ Rs and $R_c = 0.134$ Rs, so $\delta = 0.68$, which when substituted in Eq. (2) with $\gamma = 4/3$ gives $M = 1.56$. We use $R_{sh}$ measured at 05:38:54 UT and 05:40:54 UT, to get the local shock speed at 05:39:54 UT as $V_{sh} \sim 721$ km/s, which in turn gives $V_A = V_{sh}/M = 462$ km/s. Putting $n = 5.1 \times 10^7 \text{cm}^{-3}$ and $V_A = 462$ km/s in Eq. (3), we get $B = 1.51$ G corresponding to a heliocentric distance of 1.35 Rs. Dulk and McLean (1978) derived the radial dependence,

$$B(r) = 0.5(r-1)^{-1.5}, \quad \quad \quad \quad \quad \quad \quad \quad \quad \quad \quad \quad \quad \quad \quad \quad \quad \quad \quad \quad \quad (4)$$

where r is the heliocentric distance. At r = 1.35 Rs, this formula gives B = 2.4 G, which is ~60% higher than our value.

Table 1 shows the derived magnetic field values for the times during which the shock and radio measurements are available. The magnetic field also declines steadily within the SDO/AIA field of view. Note that the flux rope was observed before the shock formation and the shock left the SDO/AIA field of view before the flux rope did. The Mach number was determined assuming $\gamma$ = 4/3. The numbers in parentheses correspond to $\gamma$ = 5/3. The Mach number steadily declines from 3.72 in the beginning to 1.49 just before the shock left the SDO FOV. The ambient Alfvén speed steadily increases reaching a maximum of 462 km/s and the declines by a few percent in the next two measurements. The Alfven speed decline directly reflects the decrease in the local shock speed obtained from three consecutive height – time measurements of the shock (except for the first and last measurements). It is possible that the shock indeed begins to weaken around this time, but we do not have sufficient information because of the limited SDO FOV.

The numbers shown in Table 1 are shown graphically in Fig.5. The Alfvén speed of the corona in Fig.5a is obtained using the "local shock speed" method described above. In Fig. 5b, the shock speeds used are from the quadratic fit shown in Fig. 3a. The slight decline in Alfven speed is likely to be a local fluctuation, which is smoothed out by the quadratic fit. The Mach number declines through the SDO/AIA field of view in both cases because its derivation does not depend on the shock speed. The lower $\gamma$ value results in higher Alfvén Mach number. The Alfvén speed on the other hand is lower for the higher value of $\gamma$ because for a given shock speed, the Alfvén speed is inversely proportional to the Mach number. Weakening of the shock is evident even when the shock speed is increasing because of the rapid increase in the ambient Alfvén speed. It appears that the present observations correspond to the increasing leg of the Alfvén speed close to the coronal base (Mann et al., 1999; Gopalswamy et al., 2001; Mann et al., 2003). Gopalswamy et al. (2001) found that the Alfvén speed starts increasing around a heliocentric distance of 1.4 Rs. In the present case, the increase seems to have started a bit closer to the Sun. As noted before, the shock seems to have dissipated by the time the CME reached the LASCO/C2 FOV.

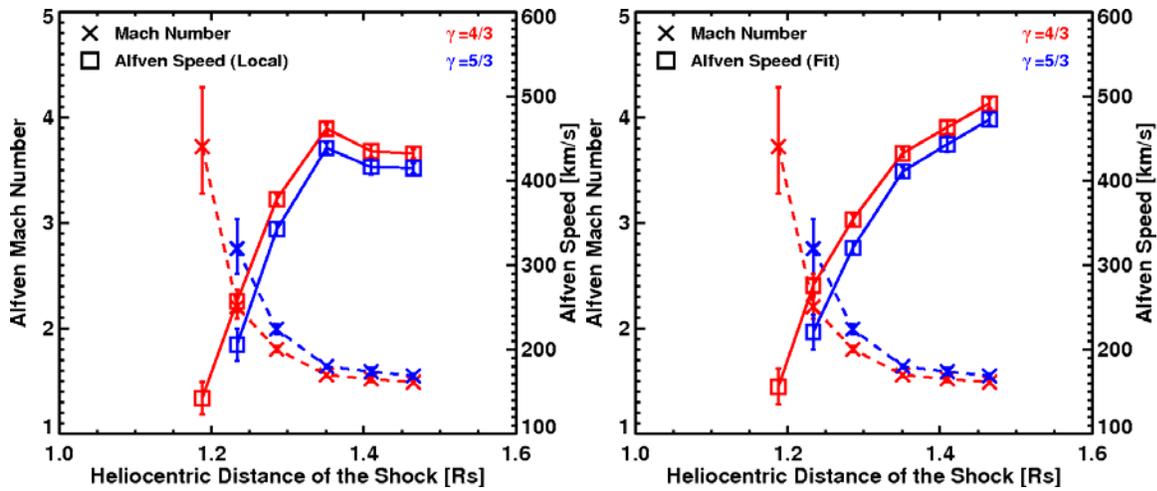

Figure 5. (a) The derived Mach number and Alfvén speed for γ =4/3 and 5/3. (b) Same as in (a) except that the Alfvén speed is obtained using a shock speed derived from a second order fit to the height-time plot.

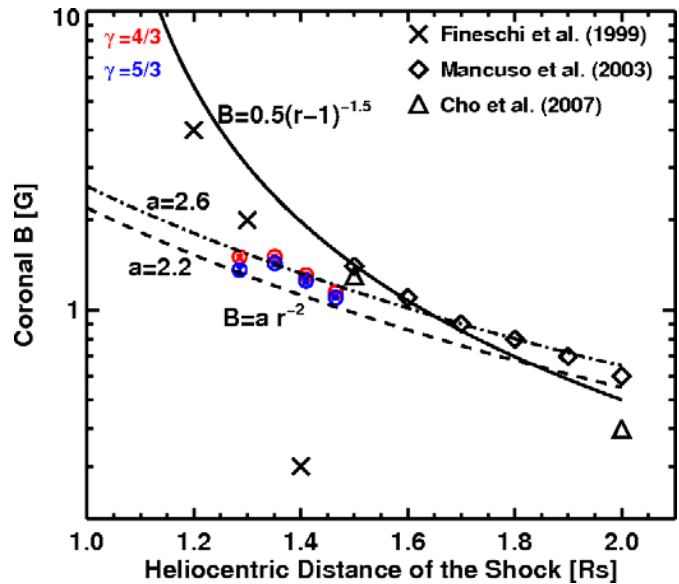

Figure 6. The derived coronal magnetic field values for γ =4/3 and 5/3 using the plasma densities inferred from the lower sideband of the type II radio burst. The Dulk and McLean (1978) empirical relation for the magnetic field above active region [$B(r) = 0.5(r-1)^{-1.5}$] corona and a quiet-Sun magnetic field model $B(r) = a/r^2$ with a = 2.2 and 2.6 are also shown. Magnetic field values from other techniques are shown for comparison.

Figure 6 shows the derived magnetic fields from the shock measurements for γ =4/3 and 5/3 at four instances when the upstream plasma density is available from the radio observations. Our magnetic field value at the lowest height is not reliable because the band splitting is not clear (see Fig.4). Also shown are the Dulk and McLean (1978) empirical relation for the magnetic field above active region corona and a quiet-Sun magnetic field model $B(r) = a/r^2$ with two values of the constant a: 2.2 and 2.6. Our values are closest to the quiet-Sun curve with a=2.6. The model curves show that the active region blends with the quiet corona around 1.4 Rs (see also Gopalswamy et al., 2001); in the present case, it seems to have happened at a lower heliocentric distance. The active region was very small and had an area of only 10 msh (http://kukui.ifa.hawaii.edu/ARMaps/Archive/2010/20100613.1632_armap.png). Therefore, it is likely that the quiet corona became dominant at lower altitudes than expected. Magnetic field values from other techniques are shown for comparison: Fineschi et al. (1999), Mancuso et al. (2003), and Cho et al. (2007). These values are in general agreement with our vales, except for the Fineschi et al. (1999) data point at 1.4 Rs.

**2.4 Shock Compression Ratio**

Finally, we check the consistency of our analysis from the compression ratio (downstream density $n_2$ to upstream density $n_1$) obtained from band splitting to the theoretical value. From the dynamic spectrum, $n_2/n_1$ is simply $(f_{p2}/f_{p1})^2 = (f_2/f_1)^2$. For the four instances in Fig.4 we get the compression ratio as 1.42, 1.60, 1.53, and 1.51. The first of these values is not very accurate because the split bands are not very well defined at this time. Assuming that the shock is quasi-perpendicular at these low heights, we can use the simplified formula (Draine & McKee, 1993),

$$n_2/n_1 = 2(\gamma+1)/\{D + [D^2+4(\gamma+1)(2-\gamma)M^{-2}]^{1/2}\}, \dots\dots\dots(5)$$

where $D = (\gamma-1) + (2/M_s^2 + \gamma/M^2)$ and $M_s = V_{sh}/C_s$ is the sonic Mach number. For a 2 MK corona, the sound speed $C_s \sim 128$ km/s, so $M_s \sim 5$ for the measured shock speeds. Ma et al. (2011) estimated an upstream temperature of 1.8 MK, similar to the value assumed here. Substituting for M and $M_s$ and taking $\gamma=4/3$ in Eq. (5), we get $n_2/n_1$ consistent with the values derived from the radio dynamic spectrum (see Table 2). The first value of $n_2/n_1$ shows the largest deviation, but the radio measurements are not very accurate for this time. The compression ratio and magnetic field strength obtained by Ma et al. (2011) at one instance (05:40 UT) are consistent with our values. Note that Ma et al. (2011) did not use the standoff distance technique to get the magnetic field. Table 2 also shows the plasma beta ($\beta = C_s^2/V_A^2$) using $C_s = 128$ km/s and $V_A$ from Table 1. The values are consistent with a low beta corona as expected.

## 3. Discussion

The primary finding of this paper is that the shock structure surrounding the CME flux rope near the Sun can be used to infer the coronal magnetic field strength. The shock standoff distance and the radius of curvature of the flux rope are directly related to the Alfvénic Mach number of the shock and the adiabatic index. From the measured shock speed, we obtained the Alfvén speed assuming the adiabatic exponent. From the emission frequency of the type II radio burst produced by the shock, we get the upstream plasma density and combine it with the Alfvén speed to get the coronal magnetic field strength. The magnetic field strength obtained is consistent with other estimates from different techniques. Given the paucity of coronal magnetic field measurements, this technique will prove to be very useful in obtaining the field strengths in the coronal region where energetic events originate. Acceleration of energetic particles by CME-driven shocks occurs around or above the heliocentric distances considered here, and hence our results provide useful constraints on the magnetic field strength involved in shock acceleration theories. The magnetic field measurement is also important in providing ground truth to the extrapolation techniques commonly employed in obtaining coronal magnetic fields from the photospheric or chromospheric magnetograms. CME measurements have become routine and the shock structure is readily discerned from white light and EUV observations, so this technique can be used whenever a CME drives a shock. Normally CME observations are used in deriving the properties of eruptive events, but here we have used them to derive the properties of the ambient medium through which the CME-driven shock propagates.

In the analysis we have tacitly assumed that the radio emission comes from near the nose of the shock. We have no imaging observation at radio wavelengths, so we cannot justify this assumption. However, at such low heights in the corona, we do expect the shock to be quasiperpendicular at most locations, so the nose region certainly is favored because of the highest shock strength. If the radio emission originates from the flanks of the shock, then the estimated magnetic field corresponds to a slightly lower height. Note that the overall position angle extent is only ~33°, so when the nose is at 1.4 Rs, the flank 15° away from the nose is at a height of only 1.35 Rs. By assuming that the radio source is located at the nose, we are making an error of ~4% in the height at which the field estimate is made. It is also possible that the radio emission comes from an extended region around the nose where the Mach number is significantly greater than 1, so the height of the shock nose is a representative height. It must be pointed out that the upstream density can be obtained by other means such as spectroscopic observations and coronal brightness measurements (see e.g., Bemporad and Mancuso, 2010). In this respect, our method of obtaining the Alfvén speed is robust and can utilize measurements from many different instruments/techniques to obtain the magnetic field.

The derived Alfvén speed in the low corona is consistent with previous estimates based on models of density and magnetic field in the corona (Mann et al., 1999; Gopalswamy et al., 2001; Mann et al., 2003; Warmuth and Mann 2005). In particular, Gopalswamy et al. (2001) estimated that the minimum in the fast mode speed occurs around 1.4 Rs and could be as low as 230 km/s and the speed increases at larger heights. The Alfvén speed shown in Fig. 5 is consistent with this estimate. The magnetic scale height is much larger than the heliocentric distance range over which we made the magnetic field measurements. For example, an inverse-square dependence of the magnetic field strength gives a scale height of ~2.8 Rs at 1.4 Rs. Our magnetic field measurement corresponds to a distance range of only 0.3 Rs around 1.4 Rs. Thus, it is difficult to obtain the radial profile of the magnetic field. Nevertheless, we see a slight declining trend in the field strength within this range. Gopalswamy and Yashiro (2011) obtained the radial profile of the magnetic field in the outer corona from the standoff distance technique and found that the profile is flatter than the inverse-square dependence. We are in the process of investigating the reasons, one of which could be the usage of the minor radius of the flux rope, while the major radius could be deciding the standoff distance. This difference does not seem to matter very close to the sun, where the whole eruption appears spherical.

## 4. Conclusions

Based on the geometrical relationship between the CME flux rope and the shock driven by it discerned from EUV images obtained by SDO/AIA, we were able to determine the Alfvénic Mach number in the corona at heliocentric distance range 1.2 to 1.5 Rs. The Mach number is in the range 3.7 to 1.5 (assuming $\gamma = 4/3$) indicating shock weakening within the field of view of the imager. From the measured shock speed and the derived Mach number, we found the Alfvén speed to increase from ~140 km/s to 460 km/s over the distance range in question. By deriving the upstream plasma density from the emission frequency of type II radio burst, we were able to derive the coronal magnetic field to be in the range 1.5 to 1.3 G over a restricted distance range (1.3 Rs to 1.5 Rs). The derived magnetic field values are consistent with other estimates in a similar distance range, thus the EUV imagers and coronagraphs can be used as coronal magnetometers. The shock compression ratio determined from the band splitting of type II radio burst is also consistent with that derived from shock theory.

Work supported by NASA's LWS program.

Figure Captions

Figure 1. Snapshots of the CME from SDO/AIA 193 Å difference images. A small section of the images from the southwest corner is shown (the image scale is indicated in the first frame). The arrow points to the early stage of the CME. The scale of the images is also indicated on the first frame in terms of a solar radius (Rs). The optical limb is also drawn. A full cadence (12 s) movie of the event is included as an electronic supplement.

Figure 2. SDO/AIA difference image at 193 Å showing the flux rope and the shock structure surrounding it. The heliocentric distances of the shock ($R_{sh}$, marked by the '+' symbol) and the flux rope ($R_{fl}$) are shown. The red crosses are the points on the flux rope used for fitting the circle. The blue cross marks the center of the fitted circle. The radius of the circle fitted to the flux rope is the radius of curvature ($R_c$) of the flux rope.

Figure 3. (a) height – time plot of the shock and flux rope with the quadratic fit to the measurements. All heights refer to the Sun center. One-minute difference images were used. The shock first appeared only at 05:36:54 UT, which coincided with the onset of the metric type II burst. The error bars on the heights give standard deviations of four independent measurements. The equations of the fitted curves are: $R_{sh}(t) = 3.7 \times 10^{-7} t^2 + 5.6 \times 10^{-4} t + 0.9$ and $R_{fl}(t) = 6.2 \times 10^{-7} t^2 - 1.1 \times 10^{-4} t + 1.1$, where $t$ is the time in seconds from 05:30:54 UT. (b) Time evolution of the flux rope radius (also known as minor radius) obtained by fitting a circle to the flux rope cross section. The error bars represent the errors in fitting the circle (the maximum error is ±0.01 Rs with most points having a much lower error). The solid curve is the third order polynomial fit to the data points, whose equation is $R_c(t) = -6.7 \times 10^{-10} t^3 + 9.7 \times 10^{-7} t^2 - 0.2 \times 10^{-4} t + 0.06$ with $t$ in seconds from 05:30:54 UT.

Figure 4. Radio dynamic spectrum from the Hiraiso Radio Spectrograph (HiRAS) showing the type II burst starting from ~05:37 UT until about 05:47 UT beyond which it is masked by the interference in the 50 – 60 MHz frequency range. The band splitting is clear between 05:39:30 and 05:43:30 UT.

Figure 5. (a) The derived Mach number and Alfvén speed for $\gamma =4/3$ and $5/3$. The Alfven speed was derived using local shock speed obtained from three consecutive height – time measurements. (b) Same as in (a) except that the Alfvén speed is obtained using a shock speed derived from a second order fit to the height-time plot (see Fig. 3a).

Figure 6. The derived coronal magnetic field values for $\gamma =4/3$ and $5/3$ using the plasma densities inferred from the lower sideband of the type II radio burst. The Dulk and McLean (1978) empirical relation for the magnetic field above active region [$B(r) = 0.5(r-1)^{-1.5}$] corona and a quiet-Sun magnetic field model $B(r) = a/r^2$ with $a = 2.2$ and $2.6$ are also shown. Magnetic field values from other techniques are shown for comparison.

Table 1. Shock and flux rope measurements and the derived properties of the corona for γ = 4/3 (5/3)

| Time UT | $R_{sh}$ Rs | $R_{fl}$ Rs | ΔR Rs | $R_c$ Rs | δ | $V_{sh}$ km/s | $f_{p1}$ MHz | $n_1$ (x$10^7$) cm$^{-3}$ | M | $V_A$ km/s | B G |
|---|---|---|---|---|---|---|---|---|---|---|---|
| 05:34:54 | ---- | 1.15 | ---- | ---- | ---- | ---- | ---- | ---- | ---- | ---- | ---- |
| 05:35:54 | ---- | 1.16 | ---- | ---- | --- | ---- | ---- | ---- | ---- | ---- | ---- |
| 05:36:54 | 1.19 | 1.17 | 0.014 | 0.080 | 0.18 | 531 | ---- | ---- | 3.72 (----) | 143 (----) | ---- |
| 05:37:54 | 1.23 | 1.20 | 0.037 | 0.115 | 0.32 | 568 | ---- | ---- | 2.21 (2.76) | 257 (206) | ---- |
| 05:38:54 | 1.29 | 1.23 | 0.060 | 0.126 | 0.47 | 682 | 78 | 7.5 | 1.80 (1.99) | 378 (342) | 1.50 (1.36) |
| 05:39:54 | 1.35 | 1.26 | 0.092 | 0.136 | 0.68 | 721 | 64 | 5.1 | 1.56 (1.64) | 462 (439) | 1.51 (1.43) |
| 05:40:54 | 1.41 | 1.30 | 0.108 | 0149 | 0.73 | 663 | 59 | 4.3 | 1.52 (1.59) | 435 (416) | 1.31 (1.25) |
| 05:41:54 | 1.47 | 1.34 | 0.126 | 0.162 | 0.77 | 644 | 52 | 3.3 | 1.49 (1.55) | 432 (415) | 1.15 (1.10) |
| 05:42:54 | ---- | 1.38 | ---- | ---- | ---- | ---- | 45 | 2.5 | ---- | ---- | ---- |

Table 2. Shock properties from theory and measurements for γ=4/3 (5/3) and sound speed = 128 km/s

| Time UT | $R_{sh}$ Rs | $V_{sh}$ km/s | M | $M_s$ | $n_2/n_1$ (radio) | $n_2/n_1$ (theory) | β |
|---|---|---|---|---|---|---|---|
| 05:38:54 | 1.29 | 682 | 1.80 (1.99) | 5.33 | 1.42 | 1.93 (2.01) | 0.11 (0.14) |
| 05:39:54 | 1.35 | 721 | 1.56 (1.64) | 5.63 | 1.60 | 1.67 (1.71) | 0.08 (0.09) |
| **05:40:54** | 1.41 | 663 | 1.52 (1.59) | 5.18 | 1.53 | 1.61 (1.65) | 0.09 (0.09) |
| **05:41:54** | 1.47 | 644 | 1.49 (1.55) | 5.03 | 1.51 | 1.57 (1.61) | 0.09 (0.10) |